\documentclass[aps,pra,twocolumn,superscriptaddress,amsmath,amssymb,longbibliography]{revtex4-1}
\usepackage[english,american,UKenglish]{babel}
\usepackage{graphicx}
\usepackage{hyperref}
\usepackage{array}
\usepackage{bbding} 
\usepackage{ wasysym }
\usepackage{amsmath,amsfonts,amssymb,bm}
\usepackage{braket}
\usepackage[ansinew]{inputenc}
\usepackage{color}
\usepackage{calc}
\usepackage[normalem]{ulem}
\usepackage{natbib}
\usepackage{bibentry}
\usepackage{pbox}

\begin{document}
\title{Atomic-scale Stark-shift spectroscopy and microscopy of organic molecules}

\author{Xabier Arrieta}
\affiliation{Centro de F\'isica de Materiales (CFM-MPC), CSIC-EHU, Donostia-San Sebasti\'{a}n 20018, Spain.}
\affiliation{Department of Electricity and Electronics,\unpenalty~FCT-ZTF,\unpenalty~EHU,\unpenalty~Leioa 48940,\unpenalty~Spain.}

\author{Sofia Canola$^\dagger$}
\affiliation{Institute of Physics,\unpenalty~Czech Academy of Sciences,\unpenalty~Cukrovarnick\'{a} 10,\unpenalty~Prague,\unpenalty~16200,\unpenalty~Czech Republic.}

\author{Ruben Esteban}
\affiliation{Centro de F\'isica de Materiales (CFM-MPC), CSIC-EHU, Donostia-San Sebasti\'{a}n 20018, Spain.}
\affiliation{Donostia International Physics Center,\unpenalty~Donostia-San Sebasti\'{a}n 20018,\unpenalty~Spain.}

\author{Javier Aizpurua}
\affiliation{Donostia International Physics Center,\unpenalty~Donostia-San Sebasti\'{a}n 20018,\unpenalty~Spain.}
\affiliation{Department of Electricity and Electronics,\unpenalty~FCT-ZTF,\unpenalty~EHU,\unpenalty~Leioa 48940,\unpenalty~Spain.}
\affiliation{IKERBASQUE, Basque Foundation for Science,\unpenalty~Euskadi Plaza 5,\unpenalty~Bilbao 48009,\unpenalty~Spain.}

\author{Tom\'a\v{s} Neuman}
\affiliation{Institute of Physics,\unpenalty~Czech Academy of Sciences,\unpenalty~Cukrovarnick\'{a} 10,\unpenalty~Prague,\unpenalty~16200,\unpenalty~Czech Republic.}

\begin{abstract}
In conventional optical Stark-shift spectroscopy, molecules are exposed to spatially homogeneous static electric fields that shift the energies of their spectral lines. These shifts are attributed to the molecular electronic properties, such as variation of dipolar moment and polarizability of the molecule associated with photo(de)excitation. 
In realistic environments containing structural defects and nanoscale heterogeneities, however, molecules experience internal electric fields that vary strongly on the molecular scale, rendering the standard Stark selection rules inapplicable.
Here we develop an extended theory of atomic-scale Stark shift, addressing such scenarios. Specifically, we present a detailed theoretical analysis of an experimentally relevant configuration where the atomically sharp tip of a light-assisted scanning tunneling microscope is used to controllably apply inhomogeneous electrostatic fields to representative molecular dyes spanning several molecular families. We decompose the total Stark shift into linear and quadratic contributions and show that they contain different information about the molecular properties. 
Concretely, spatial variations of the linear Stark shift as the tip scans across the molecule enable subnanometric mapping of the charge redistribution between ground and excited electronic states, with high sensitivity to molecular composition and chemical functionalization. The quadratic Stark contribution, in contrast, reflects changes in the conventional dipolar polarizability upon excitation. Together, these results establish nanoscale Stark-shift spectroscopy as a powerful tool for resolving excited-state charge dynamics in single molecules under realistic, strongly inhomogeneous electric fields.


\end{abstract}

\maketitle

\section{Introduction}
\begin{figure*}[t]
    \centering
    \includegraphics[width=\linewidth]{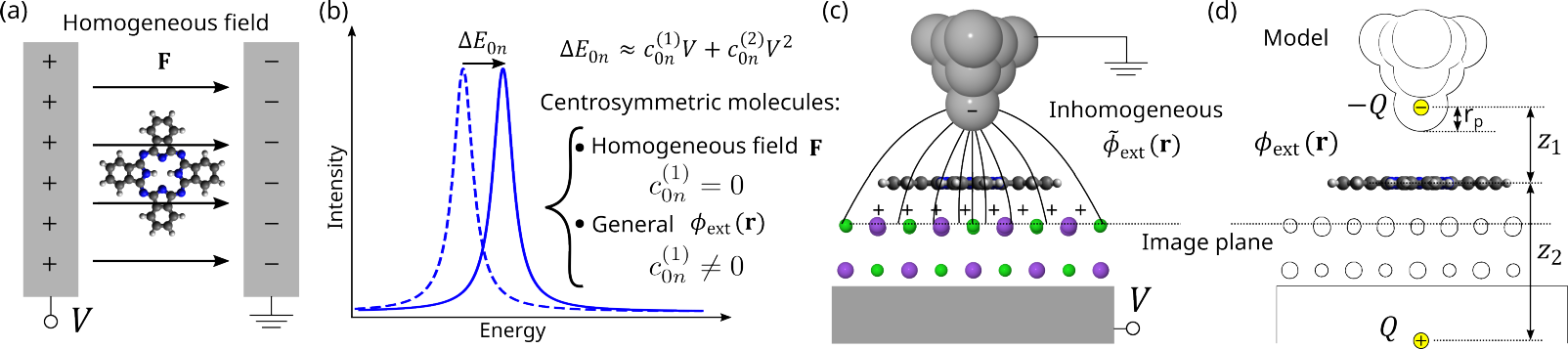}
    \caption{Illustration of Stark shifts in homogeneous and inhomogeneous fields. (a) Schematic of the standard Stark effect under a homogeneous electric field. (b) Stark shift of a molecular emission line induced by an external electric field. For centrosymmetric molecules, only quadratic Stark shift is expected under a uniform electric field $\mathbf{F}$. The application of a general inhomogeneously distributed potential $\phi_{\rm ext}$ allows even centrosymmetric molecules to exhibit both linear and quadratic Stark shifts. See the text for details. (c) Schematic of the STM setup, which produces an inhomogeneous electric field in the gap by applying a potential difference between the tip and substrate. 
    (d) Simplified model of the STM configuration, represented by two point charges (the tip charge $-Q$ and its image $Q$) that mimic the strong field localization. The point charges are positioned at a distance from the molecule $z_1=1.3\;\mathrm{nm}$ and $z_2=1.9\;\mathrm{nm}$, and an applied $V=$1V corresponds to a charge magnitude of $Q$ = 0.425e, where e is the elementary charge.}
\label{fig:main_scheme}
\end{figure*}

The DC Stark effect \cite{Stark1914} manifests as a variation of energies associated with optical emission lines of atoms and molecules when an external electrostatic field is applied. This effect has been exploited in spectroscopic techniques \cite{Bublitz1997, Boxer2009,Macfarlane2007,Malley1968,Agranovich1986,Gerblinger1987,Wild1992,Purchase1999} where usually a homogeneous electrostatic field $\mathbf{F}$ is applied (as schematically shown in  Fig.\,\ref{fig:main_scheme}a) to probe the electronic structure of systems by inspecting the changes in the molecular dipolar moment and polarizability upon optical (de)excitation. Despite its success in, e.g., revealing the nature of charge transfer excited states \cite{Boxer2009,Oh1991,Silverman2008}, actively tuning the emission energy of molecules or defects \cite{Nikolay2019, Duquennoy2024,Toninelli2021,Moradi2019}, performing Stark-shift microscopy using a tip of a near-field scanning optical microscope \cite{Karotke2006, Moerner1994}, or using emitters as probes of local electrostatic fields \cite{Friedrich2024}, the information obtained by this technique is more than than a challenge to interpret in scenarios where strongly inhomogeneous fields are at play. Such situation often occurs when the molecular environemnt contains defects, interfaces or other nanoscale inhomogeneities containing localized charges. In such situations chromophores are exposed to internal electrostatic fields that rapidly vary on nanometer length scales, thereby modifying their optical properties. This effect, called internal Stark shift, plays an important role in photosynthesis \cite{Gottfried1991,Bailleul2010} or in processes enabling vision \cite{Honig1979,Tomasello2009}, can influence electronic properties of individual isolated molecules with internal electrostatic charges \cite{Vasilev2022}, and can even be used to actively tune molecule's emission energy by optically inducing local charges in the emitter's vicinity \cite{Colautti2020}. These local charges may influence the molecular properties differently than the macroscopically applied homogeneous field as the localized electrostatic field can interact with the full multipolar distribution of molecular charge density and is not limited to interactions with the molecule's permanent dipole or induced polarizability (Fig.\,\ref{fig:main_scheme}b). Consequently, on the atomic scale, the Stark effect is governed by different selection rules than in standard experiments with homogeneous fields.

Here, we build on recent work \cite{Roslawska2022, Imada2021} and theoretically analyze the generalized concept of Stark-shift microscopy and spectroscopy exploiting the scanning-tunneling-microscope-induced-luminescence (STML) technique\cite{Chen2010,Zhang2016,Roslawska2022,Qiu2003,Imada2016,Doppagne2018,Dolezal2019,Kaiser2019,Jiang2023,Hung2023,Zhang2017a, Luo2019,Kaiser2025,Doppagne2017,Kong2021,Vasilev2024,Cao2021,Dolezal2021b}. In STML, strongly inhomogeneous static and dynamical electric fields can be controllably applied to the sample molecules with picometer spatial precision by approaching the atomically sharp microscope tip (see the schematic in Fig.\,\ref{fig:main_scheme}c). By systematically introducing a perturbative theoretical treatment of the Stark shifts, we develop a framework to study the response of molecules subjected to strongly inhomogeneous fields and to explicitly separate the linear and quadratic contributions. Using this formalism, we analyze the Stark-shift response of different molecular structures and reveal how small changes in their atomic composition can significantly influence their Stark-shift response, which is closely linked to the pattern of intramolecular charge redistribution upon excitation. To show this, we focus on a set of representative molecules with well-characterized photophysical properties, enabling a systematic study of the Stark shift under inhomogeneous fields.

\section{RESULTS AND DISCUSSION}

\begin{figure*}[t]
    \centering
    \includegraphics[width=\linewidth]{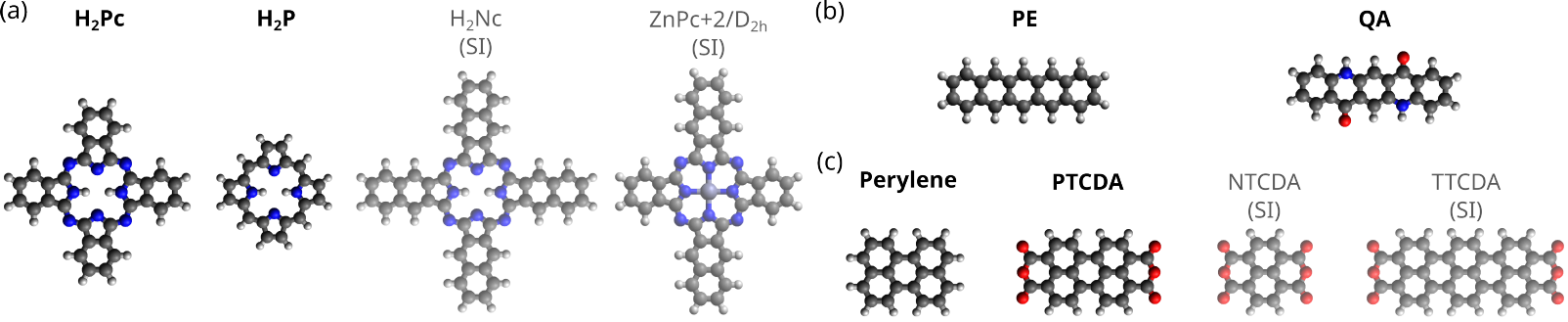}
    \caption{Chemical structure of the molecules analyzed in the work. (a) Free-base phthalocyanine (H$_2$Pc), free-base porphyrine (H$_2$P), free-base naphtalocyanine (H$_2$Nc) and Zn phthalocyanine with an asymmetric extension of the aromatic core (ZnPc+2/D$_{2\mathrm{h}}$) (b) Pentacene (PE) and quinacridone (QA). (c) Perylene and perylene-, naphthalene-, terrylene- tetracarboxylic dianhydride (PTCDA, NTCDA, and TTCDA respectively). Atom color code: H (white), C (black), N (blue), O (red), and Zn (grey). The molecules marked with (SI) are shown grayed out and are explicitly analyzed in section S3 of the supporting information.}
\label{fig:molecular_families}
\end{figure*}

To illustrate the capabilities of the framework and highlight the influence of molecular structure and atomic substitution on Stark responses, we apply our methodology to a series of representative molecules whose photophysical properties can be or have been studied experimentally with STM. 
All selected molecules are centrosymmetric in their planar geometry and, according to the standard selection rules derived within the dipolar approximation, where the molecule experiences a uniform electrostatic field $\mathbf{F}$, are expected to exhibit negligible to no linear Stark shift and only a quadratic shift proportional to changes in their polarizability. However, as we demonstrate, when the molecules are placed in electric fields strongly inhomogeneous on the scale of their geometrical size, the approximate selection rules break down and a non-zero linear component, which may be comparable to (or even dominate) the quadratic contribution, arises.
Our analysis is applied to three distinct groups of molecules, defined based on structural similarities as shown in Fig.\,\ref{fig:molecular_families} (the effects on the grayed-out molecules are discussed in the supporting information).

\textbf{Molecular Samples}. In the first group of molecules, we consider substituted phthalo- and naphthalocyanine derivatives: free base phthalocyanine (H$_2$Pc, with two central H atoms), free base naphthalocyanine (H$_2$Nc) and Zn phthalocyanine with an asymmetric extension of the aromatic core (ZnPc+2/D$_{2\mathrm{h}}$) shown in Fig.\,\ref{fig:molecular_families}a. For comparison, free-base porphyrine (H$_2$P) is also considered. All the molecules have D$_{\rm 2h}$ symmetry when planar. Calculations suggest that planarity is a good approximation for the molecular geometry when adsorbed on NaCl, as shown for H$_2$Pc \cite{Miwa2016, Vasilev2022}. 
Accordingly, we perform all calculations for this molecular family assuming planar geometries. Some of these molecules have been experimentally studied in STML \cite{Vasilev2024}, and the free-base phtalocyanine (H$_2$Pc) has even been used for imaging the tip-induced Stark shift \cite{Roslawska2022}. We can therefore use this molecule to benchmark our computational methods and demonstrate their capacity to predict experimental outcomes and provide information about the excited-state properties of the molecule.

In the second group of molecules, we consider pentacene (PE) and its isostructural analogue quinacridone (QA) (Fig.\,\ref{fig:molecular_families}b). The heteroatom substitution pattern of QA (two N atoms replacing C atoms and two O atoms replacing H atoms in a trans configuration) lowers the symmetry to the C$_{\rm 2h}$ group (with respect to the reference PE, D$_{\rm 2h}$ symmetry) but preserves the centrosymmetric character.
PE is often used in proof-of-principle STM experiments \cite{Kong2021, Kabakchiev2010, Kuhnke2017a} and QA was also recently studied with STML \cite{Jiang2023}. 
This pair of molecules provides an ideal test for comparing the Stark-shift response of two isostructural molecules that differ slightly in atomic composition, allowing us to assess how minimal chemical modifications can influence the Stark response and thereby highlight the sensitivity of atomic-scale Stark-shift microscopy. Moreover, QA bends and twists on the NaCl surface due to the strong O-Na attractive interaction (see Fig. S10 of supporting information ). In our computational analysis we consider PE in its planar geometry, while for QA we consider both the planar and non-planar geometries so as to isolate the geometrical effect on the Stark shift. 

In the third group of molecules, we consider perylene and a series of rylene derivatives of increasing length: naphthalenetetracarboxylic dianhydride (NTCDA), perylenetetracarboxylic dianhydride (PTCDA), and terrylenetetracarboxylic dianhydride (TTCDA) (Fig.\,\ref{fig:molecular_families}c). 
They have been extensively studied with STM \cite{Dori2006, Paulheim2016a, Paulheim2016, Kimura2019, Wagner2014, Wagner2015, Temirov2018, Dolezal2021b, deCamposFerreira2024, Friedrich2024, Ferreira2025} and STML \cite{Kimura2019, Dolezal2021b, deCamposFerreira2024, Ferreira2025} as they are easy to manipulate by the microscope tip and can even be attached to it to serve as molecular probes \cite{Wagner2015, Temirov2018, Esat2024}, including optically addressable ones \cite{Friedrich2024}. 
When planar, the molecules feature D$_{\rm 2h}$ symmetry.
However, on NaCl the rylene derivatives bend by anchoring the O atoms to the surface. Similarly to QA, we consider this effect separately by analyzing the rylene derivatives in both planar and non-planar geometries, allowing to isolate the impact of symmetry breaking on the Stark shift from the one driven by chemical composition. Perylene, in contrast, remains nearly planar on NaCl and is therefore treated exclusively in its planar geometry. Note that, in STM experiments, the rylene derivatives can be stabilized on insulator-covered metal surfaces at different charge states depending primarily on the substrate work function \cite{Kimura2019}. In this work, we choose to consider all molecules in the charge-neutral state, except in section S9 of the supporting information where we provide further information on the rylene derivatives in the anion state.

\textbf{Theoretical Framework}. To model the interaction between the STM-induced field and the molecule, accounting for the spatial inhomogeneities, we consider the Hamiltonian $\hat{H} = \hat{H}_0 + \hat{V}$. The unperturbed molecular Hamiltonian is written in its eigenbasis as $\hat{H}_0 = E_n^{(0)}\ket{n^{(0)}}\bra{n^{(0)}}$, with $E_n^{(0)}$ representing the energy of electronic state $n$ in the absence of an external field and $\ket{n^{(0)}}$ representing the corresponding ket state. The perturbation term $\hat V=\iiint \hat\rho({\bf r}) \phi_{\rm ext}({\bf r}-{\bf r}_{\rm t})\,{\rm d}^3 {\bf r}$ describes the interaction between the molecule and the STM-induced electrostatic potential $\phi_{\rm ext}$, where $\mathbf{r}$ is the position, $\hat{\rho}(\mathbf{r})$ is the position-dependent electron density operator of the molecule, and $\mathbf{r}_{\rm t}$ represents the tip position. At this stage one usually approximates $\hat{V}$ using the dipolar approximation. We do not perform this approximation and instead proceed considering the full 3D distribution of both the electron density and electrostatic potential. We approximate the external electrostatic potential $\phi_{\rm ext}$ as that generated by two oppositely charged point sources ($-Q$ and $Q$) located at different distances from the molecule ($z_1$ and $z_2$, respectively), as shown in Fig.\,\ref{fig:main_scheme}d. Within this model, we assume that the applied voltage drops entirely between the STM tip and the top NaCl layer, thus effectively neglecting any voltage drop within the NaCl layer (usually being about $\sim 10$\% of the total voltage \cite{Miwa2019, Repp2005a, Fatayer2018}). The top NaCl layer therefore defines the image plane of the point-charge model. The negative charge $-Q$ is placed above the molecule at the center of the atomic-scale protrusion of the STM tip, characterized by a radius $r_{\rm p}$. The corresponding image charge ($Q$) is positioned symmetrically below the molecule with respect to the image plane. The magnitude of the charges is chosen such that the resulting electrostatic potential reproduces the voltage drop between the image plane and the end of the STM tip (see section S2 of supporting information). 
The Hamiltonian $\hat H$ can be represented directly within the framework of TDDFT by including $\phi_\mathrm{ext}$ into the external potential perceived by the electrons (see Methods section). In this way, one can directly calculate the energies $\tilde E_{0n}$ of the excitations in the molecule in the presence of the perturbation and evaluate the Stark shifts as $\Delta E_{0n}=\tilde E_{0n}-E_{0n}$, where $E_{0n}=E^{(0)}_n-E^{(0)}_0$ are the excitation energies of the unperturbed system. 

In addition to this direct approach, which yields the full Stark response, we also compute the Stark shifts using perturbation theory up to second order, $\Delta E_{0n}\approx c^{(1)}_{0n} V+c^{(2)}_{0n} V^2$, with $c^{(1)}_{0n}(\mathbf{r}_{\rm t})$ and $c^{(2)}_{0n}(\mathbf{r}_{\rm t})$ only depending on the position of the tip, and study the linear $\Delta E^{(1)}_{0n}=c^{(1)}_{0n} V$ and quadratic $E^{(2)}_{0n}=c^{(2)}_{0n} V^2$ shifts independently, as different orders (linear, quadratic, and higher-order terms) generally provide different information on the molecular properties. While the linear shift reflects the intrinsic charge redistribution associated with the electronic excitation, the quadratic shift captures the ability of the electron density to reorganize under the applied field. According to first-order perturbation theory, the linear Stark shift is given by
\begin{equation}\label{eq:linear_shift}
\Delta E^{(1)}_{0n}(\mathbf{r}_{\rm t})=\iiint \Delta\rho_{0n}({\bf r}) \phi_{\rm ext}({\bf r}-{\bf r}_{\rm t})\,{\rm d}^3 {\bf r}\;,
\end{equation}
where $\Delta\rho_{0n} = \rho_n-\rho_0$ is the difference in charge densities between the excited and ground states obtained from TDDFT calculations of the isolated molecule. The quadratic Stark shift, arising from second-order perturbation theory, is expressed as (see section S1 of supporting information for a more detailed discussion):
\begin{equation}\label{eq:quadratci_shift}
    \Delta E^{(2)}_{0n}(\mathbf{r}_{\rm t})=\frac{1}{2}\iiint \Delta\delta\rho_{0n}({\bf r};\phi_{\rm ext}) \phi_{\rm ext}({\bf r}-{\bf r}_{\rm t})\,{\rm d}^3 {\bf r}\;,
\end{equation}
where $\Delta\delta \rho_{0n}=\delta\rho_n-\delta\rho_0$. The term $\delta\rho_m = \widetilde{\rho}_m-\rho_m$ represents the change in the charge density of state $m$, $\widetilde{\rho}_m$ being the charge density obtained with the influence of $\phi_{\rm ext}$, and $\rho_{m}$ being the unperturbed density. In practice, $\delta\rho_m$ is obtained by first computing $\widetilde{\rho}_m$ through TDDFT calculations that explicitly include $\phi_\mathrm{ext}$, and then subtracting $\rho_m$ obtained from the TDDFT calculations of the isolated molecule. Eqs. \eqref{eq:linear_shift} and \eqref{eq:quadratci_shift} generalize the result of the usual treatment of Stark shifts within the dipolar approximation where $\Delta E^{(1)}_{0n}\approx -\Delta \boldsymbol{\mu}_{0n}\cdot {\bf F}$ and $\Delta E^{(2)}_{0n}\approx -\frac{1}{2}{\bf F}\cdot \Delta\boldsymbol{\alpha}_{0n}\cdot {\bf F}$, with $\mathbf{F}$ being the electric field at the position of the molecule, $\Delta \boldsymbol{\mu}_{0n}=\boldsymbol{\mu}_n-\boldsymbol{\mu}_0$ being the difference between the dipolar moments of the charge density in the ground and excited states, and $\Delta\boldsymbol{\alpha}_{0n}=\boldsymbol{\alpha}_{n}-\boldsymbol{\alpha}_{0}$ being the difference between the linear polarizability tensors in the respective states \cite{Boxer2009}. 
We note that, in experiments, the linear and quadratic contributions could be extracted by measuring the Stark shift at several applied bias voltages and fitting the resulting dependence to a polynomial. Indeed, this voltage-sweep approach is equivalent to Eq.\,\eqref{eq:linear_shift} and Eq.\,\eqref{eq:quadratci_shift}, as we show in section S3 of supporting information. 

\textbf{Stark-Shift Maps}. Armed with this theoretical framework we apply it to the first group of molecules. 
In the neutral state considered here, all the molecules have a pair of singlet excited states S$_1$ and S$_2$ associated with radiative transitions from the singlet ground state that have similar excitation energy (more information about the excited states of the molecules is provided in Table S1 of supporting information). 
\begin{figure}[h!]
    \centering
    \includegraphics[width=\linewidth]{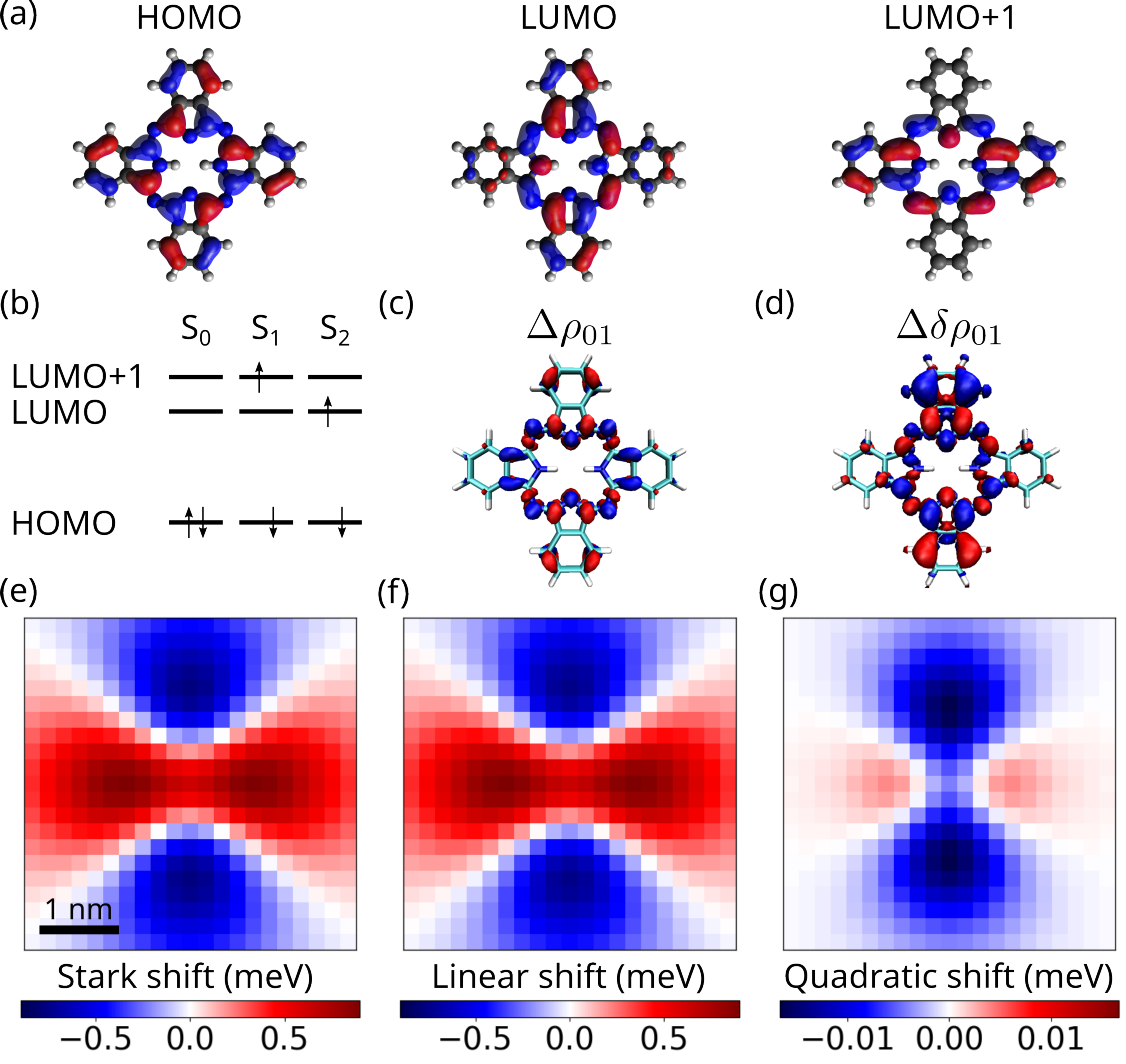}
    \caption{Electronic structure and spatially resolved Stark-shift maps for the H$_2$Pc molecule. (a) Isosurface plots of frontier molecular orbitals ($\psi$). (b) Schematic depiction of electronic configurations of states S$_0$, S$_1$, and S$_2$ of the molecule, highlighting the dominant single-electron excitation contributing to the S$_0$ to S$_1$ and S$_0$ to S$_2$ optical transitions. (c) Charge-density difference $\Delta\rho_{01}$ of H$_2$Pc, obtained from the calculation of the isolated molecule (without the STM tip), plotted as an isosurface with a value of $5\times10^{-4}e/a_0^3$, $a_0$ being the Bohr radius. (d) Difference in the field-induced charge density, $\Delta\delta\rho_{01}$, plotted as an isosurface with a value of $5\times10^{-6}e/a_0^3$, calculated by positioning  the external charges where the quadratic Stark shift (shown in panel g) is maximal. In (c) and (d) red and blue lobes represent regions of positive and negative density, respectively. (e) Total Stark shift map, (f) linear Stark shift map, and (g) quadratic Stark shift map as a function of the lateral position of the tip for S$_0$ to S$_1$ excitation of the H$_2$Pc molecule.}
    \label{fig:H2Pc}
\end{figure}

We first concentrate on the H$_2$Pc molecule as it has been studied experimentally \cite{Roslawska2022} and can be used to benchmark the calculations. The frontier molecular orbitals of H$_2$Pc (HOMO, LUMO, and LUMO+1) are shown in Fig.\,\ref{fig:H2Pc}a. 
The lowest spin-allowed excitation, S$_0$ to S$_1$, is dominated by a single-electron transition from the HOMO to the LUMO+1 (Fig.\,\ref{fig:H2Pc}b), whereas the S$_0$ to S$_2$ excitation primarily involves the HOMO to LUMO transition (analyzed in the supporting information). Because the excited state is dominated by one single-electron excitation (from $a$ to $i$), the charge density difference between the excited state $n$ and the ground state 0 can be approximated as $\Delta\rho_{0n} \approx e\left( \psi_a^2 - \psi_i^2\right)$, where $\psi_a$ and $\psi_i$ denote the spatially dependent occupied and unoccupied molecular orbitals involved in the transition (the HOMO and the LUMO+1 for the case of $\Delta\rho_{01}$ for H$_2$Pc). A detailed discussion on single-electron excitation decomposition and its use to compute the charge-density differences is provided in section S4 of supporting information. As the HOMO is evenly distributed over all arms of the molecule and the LUMO+1 is dominantly localized on one opposing pair of the molecular arms (Fig.\,\ref{fig:H2Pc}a), we expect to see significant changes of density distribution reflected in $\Delta\rho_{\rm 01}$ and, according to Eq.\,\eqref{eq:linear_shift}, also in its corresponding signatures in the (linear) Stark shift. Fig.\,\ref{fig:H2Pc}c presents the full charge-density difference $\Delta\rho_{01}$ obtained directly from TDDFT calculations of the isolated molecule, without including the STM tip (see Methods). The results confirm the expected behavior, revealing a pronounced accumulation of positive charge (i.e., a depletion of electron density, indicated by red regions) on the vertical arm, i.e., the arm perpendicular to the axis defined by the two central H atoms of H$_2$Pc, when the molecule is excited from S$_0$ to S$_1$. The quadratic shift, on the other hand, depends on the polarizability of the molecule, generalized by $\Delta\delta\rho_{\rm 01}$, which is not determined solely by the frontier molecular orbitals associated with the S$_0$ to S$_1$ excitation. To illustrate the polarization induced in the molecule by the tip, we compute $\Delta\delta\rho_{0\rm 1}$ in the presence of the external point charges placed along the vertical axis, corresponding to the tip position where the magnitude of the quadratic Stark shift is maximal, as discussed below. The resulting induced density, shown in Fig.\,\ref{fig:H2Pc}d, exhibits a clear dipolar structure oriented along the vertical axis. 

To explore how the charge-density information is reflected in the Stark-shift results, we calculate the Stark shift of the H$_2$Pc molecule as a function of the lateral position of the STM tip along the plane parallel to the one containing the atoms of the molecule, generating spatially resolved Stark-shift maps, which we analyze in terms of its linear and quadratic components. Fig.\,\ref{fig:H2Pc}e shows the total Stark shift (i.e., both linear and quadratic) obtained directly from TDDFT calculations that include the external point charges in the simulations. This provides the perturbed excitonic energies of the molecule under the influence of the external field induced by the STM. This theoretical map can be compared to experimental data previously reported in Ref. \cite{Roslawska2022} showing a good agreement upon adjustment for the applied bias. To further analyze the origin of the shifts, we show in Fig.\,\ref{fig:H2Pc}f and Fig.\,\ref{fig:H2Pc}g the linear and quadratic Stark-shift components, respectively, calculated using Eqs. \eqref{eq:linear_shift} and \eqref{eq:quadratci_shift}.
Our results show that the linear Stark shift (Fig.\,\ref{fig:H2Pc}f) for this molecule is actually the dominant component of the total shift, being an order of magnitude larger than the quadratic term for the voltage considered (Fig.\,\ref{fig:H2Pc}g). These findings justify the approach in Ref. \cite{Roslawska2022} where only the linear component was simulated. 
As dictated by Eq.\,\eqref{eq:linear_shift}, the pattern of positive and negative Stark shifts observed in the map in Figs.\,\ref{fig:H2Pc}e,f captures the main characteristics of the distribution of $\Delta\rho_{0n}$ shown in Fig.\,\ref{fig:H2Pc}c, which has dominantly quadrupolar character with a vanishing dipolar component. More concretely, the map of linear Stark shift indicates that when the molecule is excited from S$_0$ to S$_1$, the charge rearranges such that electron density accumulates on the central part of the horizontal arms (positive-shift region) while being depleted from the vertical arms (negative-shift region).  
The quadratic Stark shift, on the other hand, reflects changes in the density induced by the external point charges in the ground and excited state. These vary with the tip position, giving rise to the quadratic Stark-shift map. We find that even in the atomic-scale limit, the quadratic term is well described by the point-dipole approximation, $\Delta E^{(2)}_{0n}\approx -\frac{1}{2}{\bf F}\cdot \Delta\boldsymbol{\alpha}_{0n}\cdot {\bf F}$ (see section S3 of supporting information). 
We have verified that a similar analysis applies for H$_2$Nc and ZnPc+2/D$_{2\mathrm{h}}$ resulting in very similar Stark-shift maps (see section S5 of supporting information).

\begin{figure}[h!]
    \centering
    \includegraphics[width=\linewidth]{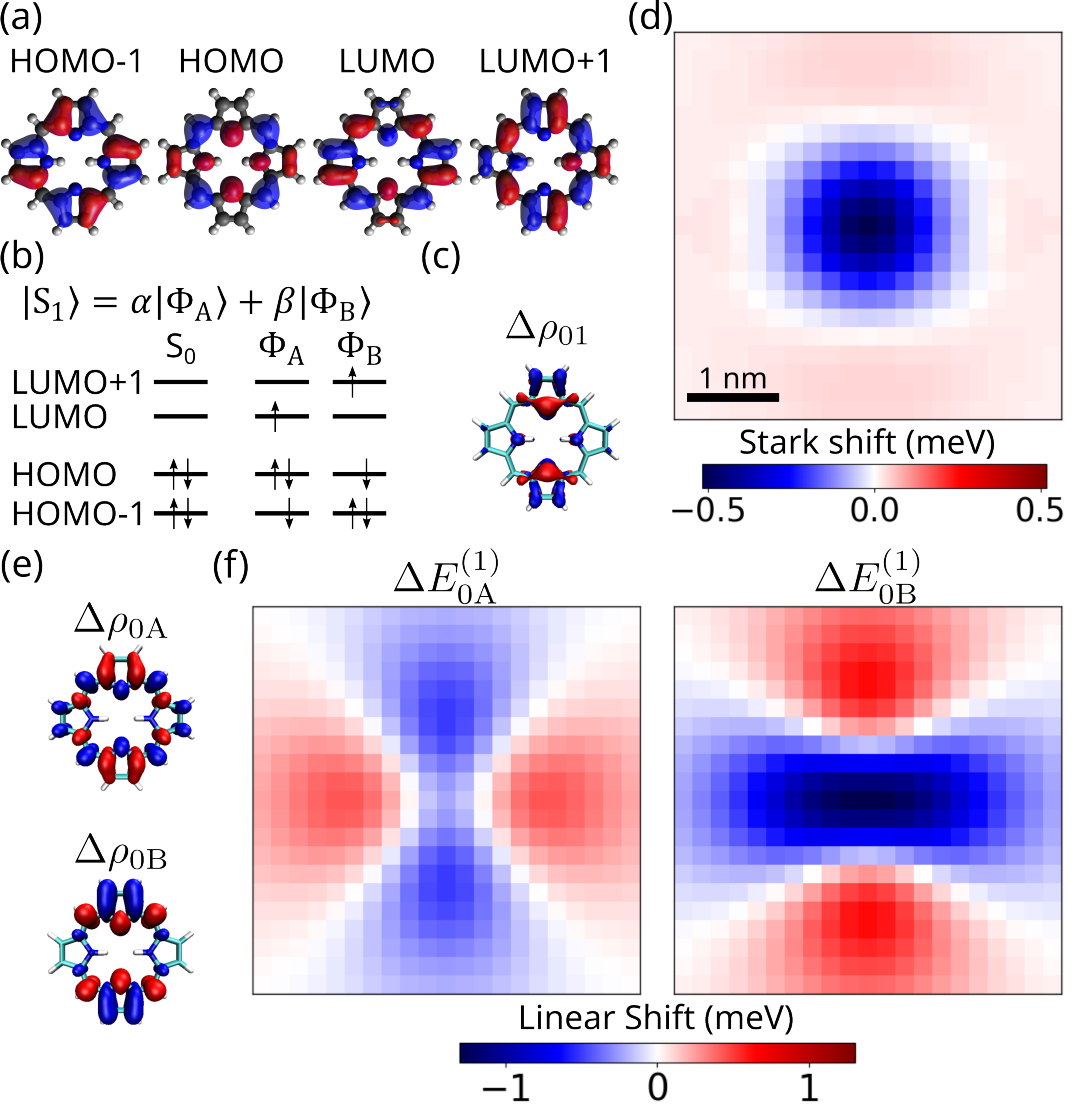}
    \caption{Electronic structure and spatially resolved Stark-shift maps for the H$_2$P molecule. (a) Isosurface plots of frontier molecular orbitals ($\psi$). (b) Schematic depiction of electronic configurations of states S$_0$ and S$_1$.  The S$_0$ to S$_1$ optical transition is expressed as a linear combination of the HOMO-1 to LUMO and the HOMO to LUMO+1 single-electron excitations, with the corresponding weighting coefficients $\alpha=0.75$ and $\beta = 0.64$ calculated from TDDFT. (c) Charge density difference plotted as an isosurface with a value of $5\times10^{-4}e/a_0^3$. (d) Total Stark-shift map as a function of the lateral position of the tip for S$_0$ to S$_1$ excitation. (e) Charge-density differences associated with the HOMO-1 to LUMO transition, $\Delta\rho_{0\mathrm{A}}$, and with the HOMO to LUMO+1 transition, $\Delta\rho_{0\mathrm{B}}$. (f) Linear Stark-shift maps corresponding to the charge-density differences from panel e: $\Delta E_{0\mathrm{A}}^{(1)}$ for the HOMO-1 to LUMO transition and $\Delta E_{0\mathrm{B}}^{(1)}$ for the HOMO to LUMO+1 transition calculated from Eq.\,\eqref{eq:linear_shift}.}
    \label{fig:H2P}
\end{figure}

The situation is different for H$_2$P as its excitations involve multiple electron-hole pairs \cite{Gouterman1961}. Namely, the lowest excited state is a superposition of the HOMO-1 to LUMO (labeled A) and the HOMO to LUMO+1 (labeled B) transitions with similar weights, Figs.\,\ref{fig:H2P}a,b. 
The resulting Stark-shift map (Fig.\,\ref{fig:H2P}d) is dominated by the linear component and it is qualitatively different from that of H$_2$Pc. In particular, in the center of the molecule, the contrast is reversed with respect to H$_2$Pc, giving rise to a negative-shift region, while the periphery exhibits a small positive shift. The calculated charge-density difference, $\Delta\rho_{01}$ (Fig.\,\ref{fig:H2P}c), shows areas of electron-density accumulation (red) situated closer to the molecular center and positive charge accumulation (blue) along the edges of the vertical molecular arms, which explains the main features observed in the Stark-shift map.
To understand the origin of this behavior, Figs.\,\ref{fig:H2P}e,f present the charge-density differences and maps of linear Stark shift associated with the two dominant single-electron transitions contributing to the S$_0$ to S$_1$ excitation. These contributions are obtained from $eC_{ia}^2\left(\psi_a^2-\psi_i^2\right)$ using the corresponding molecular orbitals and weighting coefficients ($C_{ia} = \alpha,\beta$; $a=$ HOMO-1,\,HOMO; $i =$ LUMO,\,LUMO+1), whose sum provides an approximate reconstruction of the full response, i.e., $\Delta\rho_{01} \approx \Delta\rho_{0\mathrm{A}} + \Delta\rho_{0\mathrm{B}}$ and $\Delta E_{01}^{(1)}\approx \Delta E_{0\mathrm{A}}^{(1)}+\Delta E_{0\mathrm{B}}^{(1)}$ (see section S4 of supporting information for further discussion). The results for the HOMO-1 to LUMO excitation ($\Delta\rho_{0\mathrm{A}}$ and $\Delta E_{0\mathrm{A}}^{(1)}$) indicate a charge redistribution between the pairs of molecular arms, with positive charge redistributing from the vertical molecular arms towards the horizontal arms. In contrast, the results for the HOMO to LUMO+1 excitation ($\Delta\rho_{0\mathrm{B}}$ and $\Delta E_{0\mathrm{B}}^{(1)}$) show positive charge depletion (red regions) from the molecular center towards the vertical arms. When the contributions from these two dominant single-electron excitations are combined, the total response shown in  Figs.\,\ref{fig:H2P}c,d is recovered, leaving a net charge depletion from the center towards the vertical molecular arms.
   
\begin{figure}[ht]
    \centering    \includegraphics[width=\linewidth]{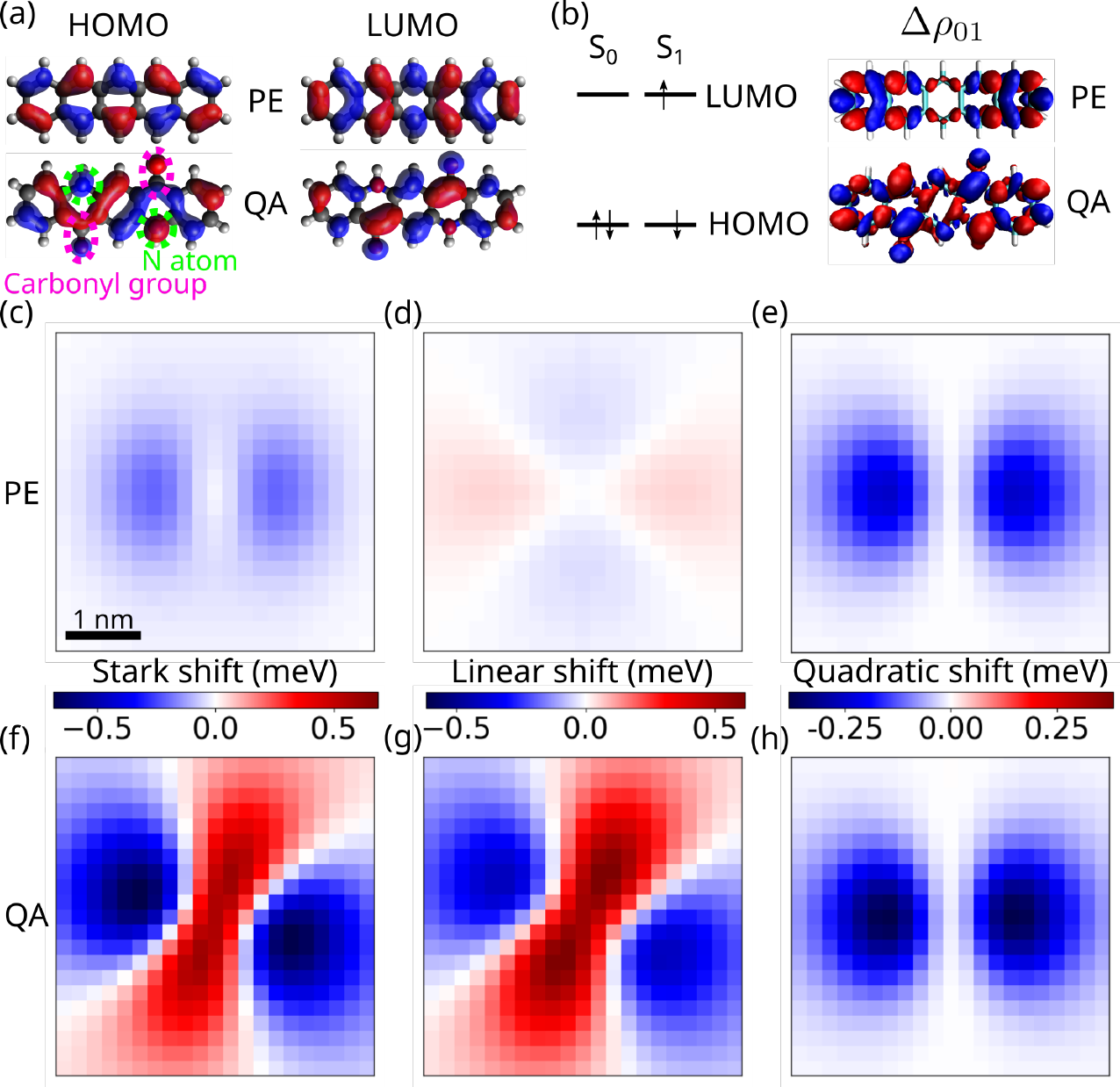}
\caption{Electronic structure and spatially resolved Stark-shift maps for PE and QA molecules. (a) Isosurface plots of frontier molecular orbitals for PE (top) and QA (bottom). (b) Schematic depiction of electronic states S$_0$ and S$_1$ for both QA and PE, highlighting that the S$_0$ to S$_1$ optical transition for both molecules is dominated by the HOMO to LUMO single-electron excitation. The corresponding isosurface plots of $\Delta\rho_{01}$ (PE top, QA bottom) are also shown at a value of $5\times10^{-4}e/a_0^3$. (c) Total Stark shift, (d) linear Stark shift, and (e) quadratic Stark shift maps as a function of the lateral position of the tip for PE. (f) Total Stark shift, (g) linear Stark shift, and (h) quadratic Stark shift maps as a function of the lateral position of the tip for QA.}
\label{fig:PEQA}
\end{figure}

The example of H$_2$Pc and H$_2$P shows that processes related to the motion of charge within the molecules as a transition is excited can be traced by measuring the Stark-shift maps. This suggests that the presence of different chemical functional groups in the molecule, which generally induce partial charges within the molecule and influence the distribution of frontier molecular orbitals, could likewise be mapped by this technique, highlighting the potential of atomic-scale Stark-shift microscopy for chemical characterization of molecular excitations. To explore this possibility, we focus on the second group of molecules consisting of PE and QA as they are a pair of isostructural molecules, differing by the presence of carbonyl groups (C=O functional groups) and N atoms in the backbone of the latter (Fig.\,\ref{fig:molecular_families}b), marked in pink and green dashed lines in Fig.\,\ref{fig:PEQA}a.
The transition from S$_0$ to the lowest excited state S$_1$ in both molecules is dominated by the HOMO to LUMO transition (Figs.\,\ref{fig:PEQA}a,b).
In PE, the HOMO and the LUMO are evenly distributed along the entire molecular structure. In contrast, in QA the presence of heteroatoms modifies the frontier molecular orbitals, where the HOMO exhibits pronounced localization over the N atom and the opposite carbonyl group and LUMO predominantly localized on the carbonyl groups. This suggests that charge redistribution upon excitation from S$_0$ to S$_1$ in QA will differ markedly from that in PE.
Indeed, in PE we see that $\Delta\rho_{01}$ features fast oscillations with only small charge redistribution 
while in QA we observe strong charge accumulation  on the heteroatoms forming an overall quadrupolar charge distribution with negative net charge around the carbonyl group and positive net charge around the N atoms. 
Following our procedure, we calculate the Stark-shift maps for both molecules and analyze the linear and quadratic components of the shift using the planar geometries of the molecules (Figs.\,\ref{fig:PEQA}c-h). For PE, the map of the total Stark shift (Fig.\,\ref{fig:PEQA}c) features two negative lobes oriented along the long axis of the molecule.
Further analysis of the shift decomposed into the linear (Fig.\,\ref{fig:PEQA}d) and quadratic (Fig.\,\ref{fig:PEQA}e) components shows that for the voltage considered the shift is dominated by the quadratic contribution caused by the differences in induced polarization of the molecule in the ground and the excited state. The linear component is minimal, as positive and negative regions of $\Delta\rho_{01}$ exhibit a highly multipolar character on the molecular scale, causing their contributions to the Stark shift to largely cancel out. This is in stark contrast with QA. For QA, the Stark-shift map (Fig.\,\ref{fig:PEQA}f) features an elongated two-lobe structure of significantly positive shift, roughly oriented along the axis defined by the O atoms, while the orthogonal direction shows two negative lobes. 
Decomposing the map into the linear (Fig.\,\ref{fig:PEQA}g) and quadratic (Fig.\,\ref{fig:PEQA}h) components, we identify that the linear shift, which dominates over the quadratic contribution, reflects the charge-redistribution pattern corresponding to an overall shift of the electrons towards the O atoms when the molecule is excited into S$_1$. Besides the linear shift, the quadratic component previously identified for PE persists in QA with a slightly larger magnitude. 

\begin{figure}[h!]
    \centering
    \includegraphics[width=\linewidth]{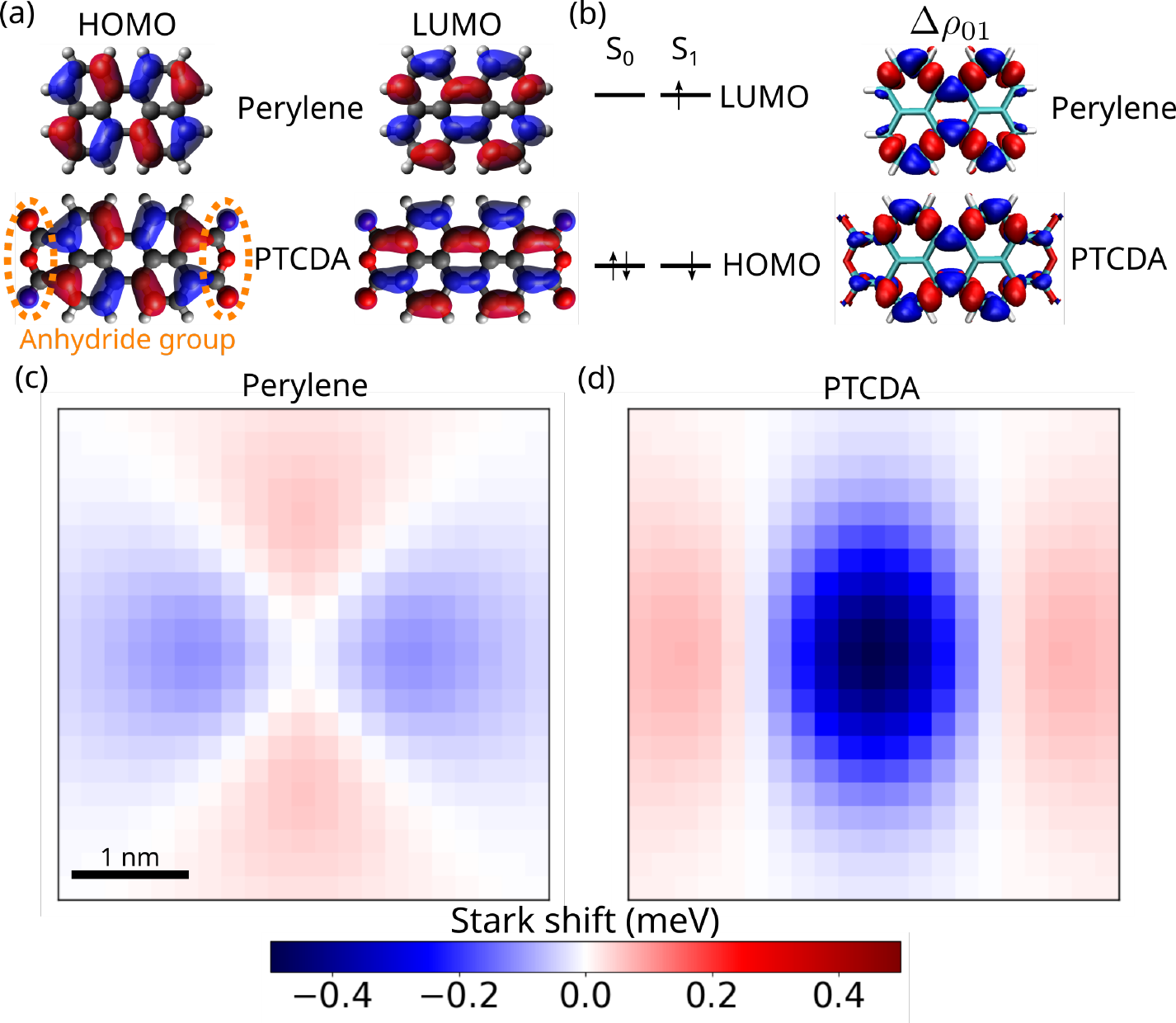}
    \caption{Electronic structure and spatially resolved Stark-shift maps for perylene and PTCDA. (a) Isosurface plots of frontier molecular orbitals for perylene (top) and PTCDA (bottom). (b) Schematic representation of the electronic configuration of states S$_0$ and S$_1$, highlighting that the S$_0$ to S$_1$ optical transition for both molecules is dominated by the HOMO to LUMO single-electron excitation. The  corresponding isosurface plots of $\Delta\rho_{01}$ (perylene top and PTCDA bottom) are also shown at a value of $5\times10^{-4}e/a_0^3$ . (c) Total Stark-shift map for perylene and (d) Total Stark-shift map for PTCDA as a function of the lateral position of the tip.}
    \label{fig:PTCDA}
\end{figure}

A similar effect can be observed in the third group of molecules when comparing perylene and PTCDA. Perylene shares the same carbon backbone as PTCDA but lacks the anhydride functional groups (-C(=O)-O-(=O)- rings at the edges of the molecule, marked by the orange dashed lines in Fig.\,\ref{fig:PTCDA}a). Comparing the results for these two molecules therefore allows us to assess the influence of the anhydride functionalities on the charge redistribution. We only show calculations done for PTCDA as it is the most commonly studied representative of n-rylenedianhydrides, but varying the length of the conjugated core (NTCDA and TTCDA) produces only minor changes in the Stark-shift maps, as shown in section S5 of supporting information.
In both PTCDA and perylene the transition from S$_0$ to the S$_1$ state is dominated by the HOMO to LUMO transition. 
The HOMO and the LUMO are similar for both perylene and PTCDA and are shown in Fig.\,\ref{fig:PTCDA}a. For the two molecules, the HOMO is localized on different C pairs compared to the LUMO; the density difference $\Delta\rho_{0_1}$ (Fig.\,\ref{fig:PTCDA}b) therefore shows prominent structure of alternating positive and negative regions along the molecular periphery for both molecules. However, $\Delta\rho_{01}$ for PTCDA shows subtle differences from that of perylene near the anhydride groups, indicating that in PTCDA the S$_1$ excitation is accompanied by a charge redistribution that is particularly pronounced near these functional groups. 
The differences in the charge-density distributions between perylene and PTCDA have an observable influence on the Stark-shift map shown in Fig.\,\ref{fig:PTCDA}c for perylene and in Fig.\,\ref{fig:PTCDA}d for PTCDA in the planar geometry, both of which are dominated by the linear component as verified in section S6 of supporting information. The map calculated for perylene shows only a small shift due to the partial cancellation of positive and negative regions in $\Delta\rho_{01}$ when computing the integral in Eq.\,\eqref{eq:linear_shift}. This confirms that, upon excitation, charge redistributes only on the scale of the interatomic distances with almost vanishing global charge redistribution on the scale of the perylene molecule, similarly to what we observed in PE.
In contrast, the PTCDA map shows a clear pattern consisting of a central spot of negative Stark shift accompanied by two less pronounced lobes on the sides near the anhydride groups. This pattern arises primarily from the global redistribution in $\Delta\rho_{01}$ towards the edges, as the fast-oscillating central part of the charge density difference largely cancels out when calculating Eq.\,\eqref{eq:linear_shift}. The Stark-shift map therefore indicates that  electrons shift towards the functional groups when the molecule is excited.
Similarly to the case of QA and PE, the differences in the Stark-shift maps between perylene and PTCDA directly reflect the presence of different functional groups, in this case anhydride groups at the edges of the PTCDA molecule.

\begin{figure}[h!]
    \centering
    \includegraphics[width=\linewidth]{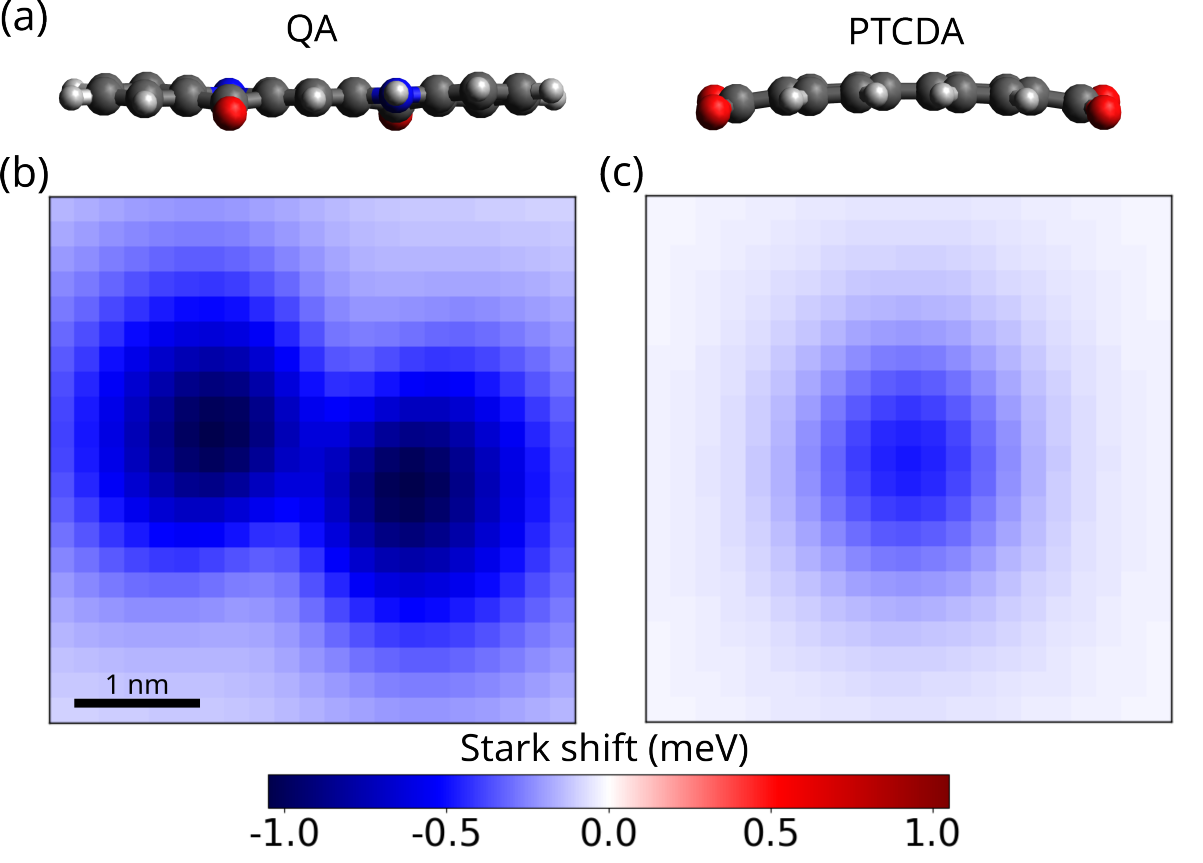}
    \caption{Spatially resolved Stark-shift maps for QA and PTCDA considering the symmetry breaking induced by the deposition of the molecules on  NaCl layers. (a) Bent geometries of QA and PTCDA obtained by optimizing the molecular structures in the presence of NaCl layers. (b) Total Stark-shift map for bent QA and (c) total Stark-shift map for bent PTCDA as a function of the lateral position of the tip.}
    \label{fig:bending}
\end{figure}

\textbf{Role of Geometry Distortion Due to NaCl Layers}. The results presented so far consider molecules in their flat geometry. When the molecules are deposited on an ionic surface (here a few NaCl monolayers), as commonly adopted in STML experiments, their geometry can be distorted by interactions with the substrate. Molecules from the first group (substituted phthalo- and naphthalocyanine derivatives), PE, and perylene are not significantly affected upon adsorption on NaCl, and therefore the planar geometry provides a good approximation to address the reconfiguration of the electronic structure. In contrast, molecules like QA and n-rylenedianhydrides experience stronger molecule-surface interactions. In these cases, the partially negatively charged O atoms of the molecules are attracted by the positively charged ions of the NaCl layers. While these forces stabilize the molecules on the substrate, they also induce geometric distortions. To account for the effects of these geometric distortions, the ground-state equilibrium geometry of QA and PTCDA under the effect of the NaCl layers is shown in Fig.\,\ref{fig:bending}a (see section S7 of supporting information for the full geometry including the NaCl layers), where the molecular structure departs from the flat centrosymmetric geometry. For both molecules, the O atoms, which are largely responsible for the shifts of charge density within the molecule, bend toward the substrate. As a result, the molecules lose their centrosymmetric geometry and acquire a net dipole moment in the out-of-plane direction, which is expected to alter the Stark-shift response of the molecules. 
This is confirmed by the calculated Stark-shift maps shown in Figs.\,\ref{fig:bending}b,c for QA and PTCDA, respectively, which display a dominant negative shift superimposed on the characteristic pattern of the flat molecule.
We discuss in section S7 of supporting information possible approaches to recover the Stark-shift features of the flat geometry from the bent results. The first approach consists of estimating the magnitude and position of the out-of-plane dipole moments induced by the molecular bending and subtracting their contributions from the Stark shift of the bent molecules. The second approach, which may be more feasible experimentally since it does not require any prior knowledge of the dipole moments, involves measuring the Stark shift of the bent molecule at different vertical positions of the STM tip and calculating the corresponding derivative. Because these strategies provide ways to recover the results for the flat molecule, the distortion of geometry, although introducing an inconvenient background, should not jeopardize the overall ability of the method to probe the charge redistribution upon excitation.

\section{Conclusions}
We have developed a generalized framework for modeling optical Stark shifts in organic molecular emitters induced by local electrostatic fields that vary rapidly on the atomic scale. By performing a series of numerical calculations on several molecular families--including phthalocyanines, acenes, quinacridone, and rylene derivatives-- we have shown that the presence of an inhomogeneous field associated with nearby charge defects in the emitter's environment may cause Stark shifts that scale both linearly and quadratically with the amplitude of the applied field. Due to the inhomogeneous character of this local field, strong linear optical Stark shifts appear even in centrosymmetric molecules, which manifests a breakdown of dipolar selection rules commonly established for situations where the molecule is exposed to an  homogeneous external electric field. 

We have revealed the relative importance of linear and quadratic contributions to the Stark shift and identified the microscopic physical mechanisms promoting them. Our calculations suggest that, by detecting these linear and quadratic shifts as a function of the lateral position of the STM tip, one could respectively recover the variations of the charge distribution and the overall static electronic polarizability between the molecule's ground and excited states. Furthermore, we have shown that chemical modifications, such as the addition of functional groups, can substantially alter the Stark shift, highlighting the sensitivity of this technique to molecular structure.
This could lead to the advent of atomic-scale Stark-shift microscopy as a tool to image charge redistribution (transfer) upon exciting organic molecules and thus probe molecular chemical structure. A particularly promising application could lie in directly imaging processes in molecular charge-transfer complexes where Stark shift microscopy could reveal the regions of charge depletion and accumulation upon photoexcitation. Beyond chemistry, atomic-scale Stark-shift spectroscopy could be employed to assess the energetic stability of molecular absorption (emission) in the presence of strongly inhomogeneous external fields, for example, those generated by nearby defects. Such an approach may contribute to the development of robust molecular (single-)photon emitters, for which the stability of their photon-emission line is key to their technological feasibility. 

\section{Methods}
All calculations were performed using time-dependent density functional theory (TD-DFT) as implemented in Gaussian 16 revision B.01 \cite{Gaussian16}. The B3LYP exchange-correlation functional \cite{Becke1993,Lee1988} together with the 6-31G* basis set was employed throughout the main text. To asses the dependence of the result on the choice of functional, additional calculations were carried out using the long-range corrected functional $\omega$B97XD \cite{Chai2008}, These  results are presented in section S8 of the supporting information and show no qualitative modifications when compared to the results discussed in the main text. 
For each isolated molecule, the ground-state geometry was first optimized. Vertical excited-state energies and charge densities were then computed at the optimized geometry. These results served as the reference for the Stark-shift analysis. To generate spatially resolved total Stark-shift maps, external point charges, which mimic the electrostatic perturbation induced by the STM tip, were introduced in the TDDFT calculation alongside the molecule. The point charges were laterally displaced over a grid above and below the molecular plane, and for each position, the Stark shift was calculated as the change in excitation energy relative to the isolated molecule. The linear Stark component was evaluated using Eq\,\eqref{eq:linear_shift}. First, the charge densities of the ground state ($\rho_0$) and excited state ($\rho_n$) of the isolated molecule were extracted from the output using the Cubegen utility of Gaussian 16. The charge density difference, $\Delta\rho_{0n} = \rho_n-\rho_0$, was then integrated over space with the external electrostatic potential $\phi_\mathrm{ext}$ generated by two point charges as shown in Fig.\,\ref{fig:main_scheme}d. The quadratic Stark component, given by Eq.\,\eqref{eq:quadratci_shift}, was obtained from the TDDFT calculations that explicitly include the external point charges, i.e., the same calculations used to determine the total Stark shift. For each lateral charge position, modified charge densities are obtained for the ground ($\widetilde{\rho}_0$) and excited state ($\widetilde{\rho}_n$). Subtracting the corresponding densities of the isolated molecule provides the induced charge densities ($\delta\rho_0$ and $\delta\rho_{n}$). The quadratic Stark shift was then calculated by integrating the difference between the induced charge densities $\Delta\delta\rho_{0n} = \delta\rho_n - \delta\rho_0$ with the external potential $\phi_\mathrm{ext}$ generated by two point charges.

\section*{Acknowledgments}
TN and SC acknowledge the Lumina Quaeruntur fellowship of the Czech Academy of Sciences and QUANTERA project MOLAR with reference PCI2024-153449 supported by the Ministry of Education, Youth and Sports (MSMT) of the Czech Republic. Computational resources were supplied by the project "e-Infrastruktura CZ" (e-INFRA CZ LM2018140) supported by the Ministry of Education, Youth and Sports of the Czech Republic. JA, RE and XA acknowledge grant PID2022-139579NB-I00 funded by MICIU/AEI/10.13039/501100011033 and by ERDF, EU, grant no. IT 1526-22 from the Department of Science, Universities and Innovation of the Basque Government, and  Elkartek project "u4smart" from the Dept. of Industry of the Basque Government.  XA acknowledges Spanish Ministerio de Ciencia, Innovaci\'on y Universidades for his PhD grant no. FPU21/02963. 

\bibliographystyle{apsrev4-1}
\bibliography{biblio.bib}

\end{document}